\documentclass[a4paper,fleqn]{cas-dc}
\usepackage[utf8]{inputenc}
\usepackage{textgreek}
\usepackage{amsmath}
\usepackage[numbers]{natbib}
\usepackage[utf8]{inputenc}
\DeclareUnicodeCharacter{22EF}{\dots}

\usepackage{soul}
\usepackage{siunitx}

\def\tsc#1{\csdef{#1}{\textsc{\lowercase{#1}}\xspace}}
\tsc{WGM}
\tsc{QE}
\tsc{EP}
\tsc{PMS}
\tsc{BEC}
\tsc{DE}

\begin{document}
\let\WriteBookmarks\relax
\def\floatpagepagefraction{1}
\def\textpagefraction{.001}
\shorttitle{Journal of Energy Storage}
\shortauthors{Laranjeira et~al.}

% Main title of the paper
\title [mode = title]{$\beta$-Irida-Graphene: A New 2D Carbon Allotrope for Sodium-Ion Battery Anodes}

\author[1]{José A. S. Laranjeira}
\affiliation[1]{
organization={Modeling and Molecular Simulation Group},
addressline={São Paulo State University (UNESP), School of Sciences}, 
city={Bauru},
postcode={17033-360}, 
state={SP},
country={Brazil}}
\cormark[1]
\cortext[cor1]{Corresponding author}
\credit{Conceptualization of this study, Methodology, Investigation, 
Formal analysis, Writing -- review \& editing, Writing -- original draft}
\author[2]{Kleuton A. L. Lima}
\affiliation[2]{
organization={Department of Applied Physics and Center for Computational Engineering and Sciences},
addressline={State University of Campinas}, 
city={Campinas},
postcode={13083-859}, 
state={SP},
country={Brazil}}
\credit{Conceptualization of this study, Methodology, Investigation, Formal analysis, Writing -- review \& editing, Writing -- original draft}

\author[1]{Nicolas F. Martins}
\credit{Conceptualization of this study, Methodology, Investigation, 
Formal analysis, Writing -- review \& editing, Writing -- original draft}
\author[3,4]{Luiz A. Ribeiro Junior}
\affiliation[3]{
organization={Institute of Physics},
addressline={University of Brasília}, 
city={Brasília },
postcode={70910‑900}, 
state={DF},
country={Brazil}}
\affiliation[4]{
organization={Computational Materials Laboratory, LCCMat, Institute of Physics},
addressline={University of Brasília}, 
city={Brasília},
postcode={70910‑900}, 
state={DF},
country={Brazil}}
\credit{Conceptualization of this study, Methodology, Investigation, 
Formal analysis, Writing -- review \& editing, Writing -- original draft}

\author[2]{Douglas S. Galvão}
\credit{Conceptualization of this study, Methodology, Investigation, 
Formal analysis, Writing -- review \& editing, Writing -- original draft}

\author[5]{{Luis A. Cabral}}
\credit{Investigation, Formal analysis, Resources, Writing -- review \& editing}
\affiliation[5]{organization={Department of Physics and Meteorology},
addressline={São Paulo State University (UNESP), School of Sciences}, 
city={Bauru},
% Uncomment if no comma needed between city and postcode
postcode={17033-360}, 
state={SP},
country={Brazil}}

\author[1]{Julio R. Sambrano}
\cormark[2]
\cortext[cor2]{Main corresponding author}
\credit{Conceptualization of this study, Methodology, Investigation, 
Formal analysis, Writing -- review \& editing, Writing -- original draft}

\begin{abstract}
The quest for sustainable and efficient energy storage has driven the exploration of sodium-ion batteries (SIBs) as promising alternatives to lithium-ion systems. However, the larger ionic radius of sodium poses intrinsic challenges such as slow diffusion and structural strain in conventional electrode materials. 
As a contribution to addressing these limitations, the $\beta$-Irida-graphene ($\beta$-IG) is herein introduced, a novel two-dimensional (2D) carbon allotrope derived from Irida-graphene, featuring a diverse polygonal lattice of 3-, 4-, 6-, 8-, and 9-membered carbon rings. Through density functional theory and \textit{ab initio} molecular dynamics simulations, $\beta$-IG demonstrated remarkable thermal, dynamical, and mechanical stability, coupled with intrinsic conductive character and efficient sodium-ion mobility (energy barriers < 0.30 eV). Furthermore, 
the adsorption of sodium ions was energetically favorable, delivering an impressive predicted specific capacity of 554.5 mAh/g. The reported findings highlight $\beta$-IG as a good potential anode candidate for next-generation SIBs, offering high-rate performance and structural robustness, and expanding the functional design space for advanced carbon-based electrode materials.
\end{abstract}

% Use if graphical abstract is present
% \begin{graphicalabstract}
% \includegraphics{figs/grabs.pdf}
% \end{graphicalabstract}

\begin{highlights}
\item $\beta$-Irida-graphene is a novel carbon monolayer with 3-, 4-, 6-, 8-, and 9-membered rings
\item The monolayer exhibits metallic behavior and thermal stability at 300~K
\item Na adsorptions resulted in strong binding energies of –2.0~eV
\item Low Na energy barriers (0.16–0.27~eV) yield high ionic mobility
\item Theoretical Na storage capacity reaches 554.5~mAh$\cdot$g$^{-1}$ with a stable OCV profile
\end{highlights}


\maketitle

\section{Introduction}

The rapid expansion of portable electronics, electric vehicles, and grid-scale energy storage systems has created a growing demand for high-performance rechargeable batteries. Among the various battery technologies explored, lithium-ion batteries (LIBs) have dominated the market for decades due to 
their high energy density and long cycle life~\cite{LIU2021102332, NZEREOGU2022100233, Manthiram2020}. However, concerns regarding the uneven geographical distribution, limited availability, and rising cost 
of lithium resources have stimulated the search for alternative energy storage technologies based on more earth-abundant elements~\cite{doi:10.1021/acsnano.9b04365, doi:10.1021/acscentsci.7b00288}.

Sodium-ion batteries (SIBs), in particular, have emerged as a viable low-cost alternative, given the natural abundance, low cost, and similar intercalation chemistry compared to lithium~\cite{Vaalma2018, REHM2025236290, doi:10.1021/acsenergylett.0c02181, doi:10.1021/acsmaterialsau.3c00049}. 
Despite these advantages, the development of efficient SIBs remains hindered by several intrinsic challenges, such as the larger ionic radius of Na$^+$, which leads to sluggish diffusion kinetics, greater structural strain, and lower specific capacities in conventional electrode 
materials~\cite{QIU2024103760, https://doi.org/10.1002/aenm.202302321, doi:10.1021/acsnano.3c02892, https://doi.org/10.1002/anie.201703772}.

To address these features, the discovery and rational design of new anode materials capable of reversibly storing sodium while maintaining structural integrity and offering high conductivity have become a critical research focus \cite{https://doi.org/10.1002/adma.202404574,https://doi.org/10.1002/chem.201402511, https://doi.org/10.1002/smll.202101137, https://doi.org/10.1002/smll.201701835}. In this context, two-dimensional (2D) carbon allotropes have received significant attention due to their low mass density, excellent electrical conductivity, high surface-to-volume ratio, and structural tunability \cite{D4NR05087H, D3CP05553A, doi:10.1021/acsanm.3c01925, 10.1063/5.0247586, 10.1063/5.0141032}. This is also due in part to the advent of graphene~\cite{doi:10.1126/science.1102896, https://doi.org/10.1002/pssb.200776208}, which created a revolution in materials science. Recent breakthroughs in the synthesis and simulation of non-honeycomb carbon lattices include biphenylene lattices~\cite{doi:10.1126/science.abg4509}, 2D fullerene layers~\cite{Hou2022}, and structures incorporating four-, five-, seven-, and eight-membered rings~\cite{Lu04032022, LIMA2025116099, 
10.1063/5.0235819, Li_2025, Jana_2022}. 

A notable development in this field is the recently proposed Irida-graphene (IG)~\cite{PEREIRAJUNIOR2023100469}, a 2D carbon structure composed solely of $sp^2$-hybridized atoms arranged in a periodic 3--6--8 ring pattern. 
IG exhibits not only thermal and dynamical stability but also a Dirac-like dispersion above the Fermi level. Computational investigations have revealed its good potential as an SIB anode, with a predicted 
capacity exceeding 1000~mAh$\cdot$g$^{-1}$, ultralow diffusion barriers ($\sim$0.09 eV), and stable voltage profiles~\cite{MARTINS2024114637}. Moreover, IG has demonstrated versatility in other energy storage applications, such as hydrogen storage, where its decoration with OLi$_3$ clusters yielded gravimetric 
capacities (10.0~wt\%) surpassing U.S. DOE targets~\cite{LARANJEIRA2025112951}.

Motivated by the numerous reports that have demonstrated the potential of IG for energy applications in the last two years \cite{TAN2024738, D4NR02669A, KAUR2024114456, XIONG2024113225, ZHANG2024118, ZHANG20241004, DUAN20241, YUAN2024114756, D4CP03381G}, it is proposed here a new structure, named $\beta$-Irida-graphene ($\beta$-IG) (Figure~\ref{fig:structure}), a topological 
derivative of IG featuring a richer assortment of polygonal motifs (3-, 4-, 6-, 8-, and 9-membered rings) embedded in a rectangular lattice. The $\beta$-IG potential as an SIB anode was systematically investigated. Employing first-principles simulations, it was demonstrated that $\beta$-IG maintains excellent thermal and mechanical stability, exhibits metallic behavior, and high Na-ion mobility, while presenting a significant storage capacity. These features pose $\beta$-IG as a promising candidate for next-generation sodium-ion battery 
anodes and expand the design space for functional 2D carbon allotropes in electrochemical applications.

\section{Computational Modeling}

To characterize the $\beta$-IG monolayer and assess its performance as an anode material for sodium-ion batteries, first-principles calculations within the density functional theory (DFT)  framework~\cite{becke2014perspective} were carried out, associated with \textit{ab initio} molecular dynamics (AIMD) simulations~\cite{marx2000ab}. These simulations employed the generalized gradient approximation (GGA) as implemented by Perdew, Burke, and Ernzerhof (PBE)~\cite{PhysRevLett.77.3865} for the exchange-correlation 
energy functional. Core-electron interactions were treated using the projector augmented wave (PAW) method~\cite{PhysRevB.50.17953}, as implemented in the Vienna \textit{ab initio} Simulation Package  (VASP)~\cite{kresse1993ab,kresse1996efficient}. The plane-wave basis set expansion of electronic wavefunctions used a kinetic energy cutoff of 520 eV, while a vacuum buffer region of 15~\AA~was used along the perpendicular ($z$-axis) direction to prevent spurious interactions between periodic (mirror) images.

Geometry optimization and projected density of states (PDOS) calculations were performed at the $\Gamma$-centered  \textbf{k}-point grids of $6 \times6 \times 1$ and $9 \times 9 \times 1$, respectively. To account for long-range dispersion interactions, Grimme’s DFT-D2 correction was applied~\cite{grimme2006semiempirical}. Structural relaxations employed the conjugate gradient algorithm, adopting convergence criteria of $1\times10^{-5}$ eV for total energy changes and maximum Hellmann-Feynman atomic forces of 0.01 eV/\AA. AIMD simulations were carried out to assess the structural and thermodynamic stability of sodium adsorption on $\beta$-IG at 300 K. The simulations employed a 
Nosé–Hoover thermostat~\cite{hoover1985canonical} (NVT ensemble), with a 1 fs time step and a total simulation duration of 5 ps. Additionally, Bader charge analyses were performed to quantify charge transfer upon sodium  adsorption~\cite{henkelman2006fast}. To determine the sodium diffusion pathways and their energetic barriers on the $\beta$-IG surface, the climbed-image nudged elastic band (CI-NEB) method~\cite{10.1063/1.1495401, doi:10.1142/9789812839664_0016} as implemented in the Quantum Espresso package~\cite{Giannozzi_2009} was used.

The energetic stability of $\beta$-IG was analyzed by its cohesive energy (\(E_{\text{coh}}\)):

\begin{equation}
E_{\text{coh}} = \frac{E_{\text{$\beta$-IG}} - \sum_i n_iE_i}{\sum_i n_i},
\end{equation}

\noindent where \(E_{\text{$\beta$-IG}}\) is the total energy of the structure, \(E_i\) represents the energy of one 
isolated C atom, and \(n_i\) is the number of atoms that compose the $\beta$-IG unit cell. The same approach 
was used to obtain the $E_{\text{coh}}$ for the other 2D structures discussed in this work.

The capacity of $\beta$-IG to adsorb Na atoms was made by calculating the adsorption energies 
($E_{\text{ads}}$), defined as:

\begin{equation}
E_{\text{ads}} = E_{(\beta\text{-IG}+\text{Na})} - E_{(\beta\text{-IG})} - E_{(\text{Na})},
\end{equation}

\noindent where $E_{(\beta\text{-IG}+\text{Na})}$ represents the total energy of sodium adsorbed on 
$\beta$-IG, $E_{(\beta\text{-IG})}$ is the energy of the pristine monolayer, and $E_{(\text{Na})}$ denotes the energy of an isolated sodium atom.

%Charge transfer characteristics between sodium atoms and the monolayer were further analyzed using the 
%charge density difference (CDD), computed as:
%\begin{equation}
%\Delta \rho = \rho_{\beta\text{-}IG+Na} - \rho_{\beta\text{-}IG} - \rho_{\text{Na}},
%\end{equation}

%\noindent where $\rho_{\beta\text{-}IG+Na}$, $\rho_{\beta\text{-}IG}$, and $\rho_{\text{Na}}$ refer 
%to electron charge densities for sodium-decorated $\beta$-IG, pristine monolayer, and isolated sodium 
%atom, respectively.

To investigate sodium-ion diffusivity, diffusion coefficients ($D$) were computed according to the 
Arrhenius relation~\cite{gomez2024tpdh}:

\begin{equation}
D(T) = L^2 \nu_0 \exp\left(-\frac{E_{\text{barr}}}{k_B T}\right),
\label{eq:arrhenius}
\end{equation}

\noindent where $L$ denotes the diffusion length, $\nu_0$ is the attempt frequency ($1\times10^{13}$~Hz, 
derived from phonon calculations), $E_{\text{barr}}$ is the diffusion barrier obtained from CI-NEB 
calculations, $k_B$ is Boltzmann's constant ($1.38\times10^{-23}$ J/K), and $T$ is the temperature.

\section{Results and Discussion}

\subsection{Structure and Stability}

\begin{figure}[htbp]
\centering
\includegraphics[width=1\linewidth]{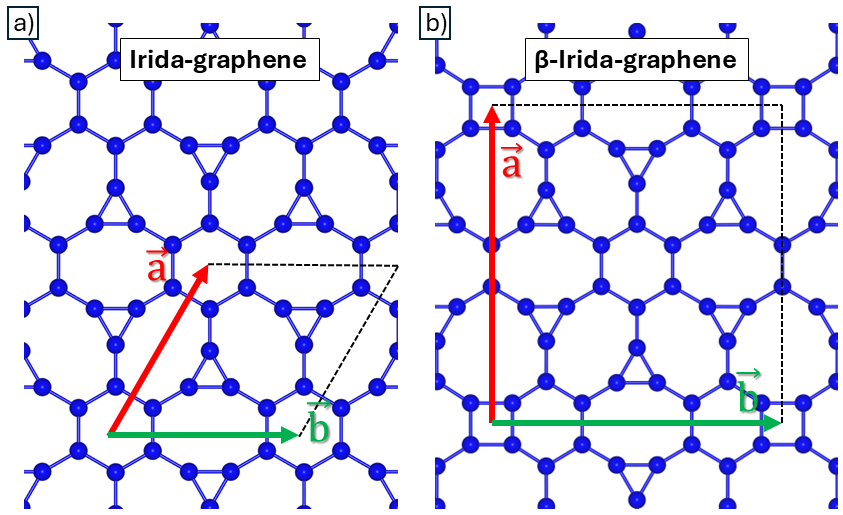}
\caption{Top views of the atomic structures and lattice vectors of (a) Irida-graphene and (b) $\beta$-Irida-graphene. While both are composed exclusively of $sp^2$-hybridized carbon atoms, $\beta$-IG introduces a richer diversity of polygonal rings (3-, 4-, 6-, 8-, and 9-membered units), arranged in a rectangular lattice with lattice vectors $\vec{a}$ (red) and $\vec{b}$ (green).}
\label{fig:structure}
\end{figure}

Figure~\ref{fig:structure} compares the atomic structures of Irida-graphene and its topological derivative, the newly proposed $\beta$-Irida-graphene ($\beta$-IG). Panel (a) illustrates the Irida-graphene 
(IG) lattice, denoted by a periodic network of 3-, 6-, and 8-membered carbon rings. Panel (b) displays the relaxed geometry of $\beta$-IG, which preserves the $sp^2$ hybridization but introduces a richer set of polygonal motifs—namely 3-, 4-, 6-, 8-, and 9-membered rings-within a rectangular lattice. The in-plane lattice parameters of $\beta$-IG are $a = \SI{10.12}{\angstrom}$ and $b = \SI{11.02}{\angstrom}$, consistent with orthorhombic symmetry (space group \textit{Pmmm}, No.~47). Structurally, $\beta$-IG can be considered as a topological 
derivative of the previously reported IG, but with a richer set of ring motifs and reduced symmetry. The cohesive energy ($E_\text{coh}$) calculated for the $\beta$-IG monolayer (-7.47 eV/atom) is in the same range as other 2D carbon allotropes. Specifically, $\beta$-IG exhibits a cohesive energy very close to that of IG (-7.50 eV/atom) and T-graphene (-7.45 eV/atom), and slightly lower than graphenylene (-7.33 eV/atom). Nevertheless, the cohesive energy remains higher than that of pristine graphene (-8.02 eV/atom).

\begin{figure}[htbp]
\centering
\includegraphics[width=0.75\linewidth]{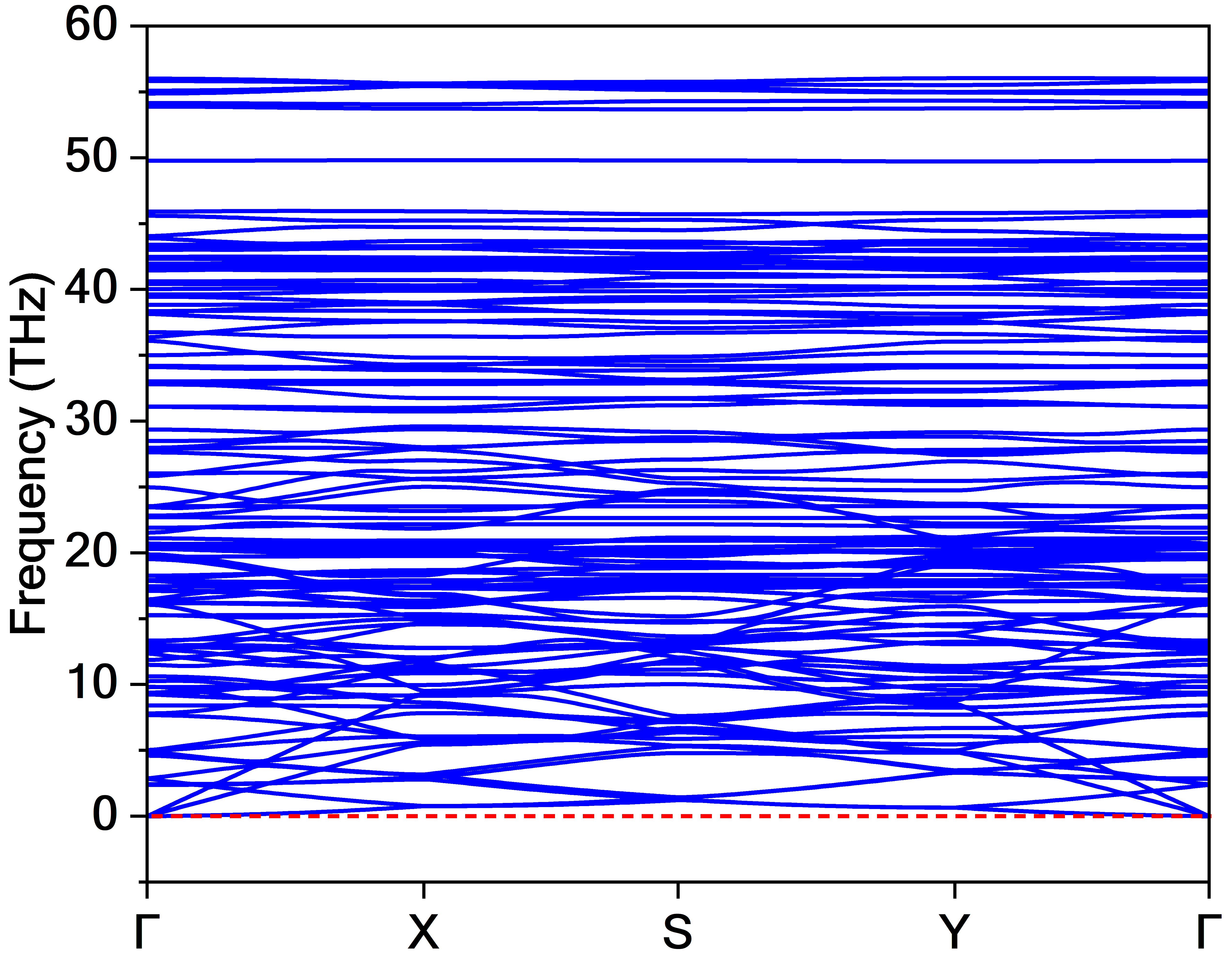}
\caption{Phonon band structure of $\beta$-Irida-graphene along the high-symmetry path $\Gamma$--X--S--Y--$\Gamma$. 
The absence of imaginary frequencies confirms the dynamical stability of the monolayer.}
\label{fig:phonon}
\end{figure}

Figure~\ref{fig:phonon} depicts the phonon dispersion curve along the high-symmetry path $\Gamma$--X--S--Y--$\Gamma$ in the Brillouin zone. The absence of imaginary frequencies throughout the whole spectrum confirms the $\beta$-IG dynamical stability.

\begin{figure}[htbp]
\centering
\includegraphics[width=1\linewidth]{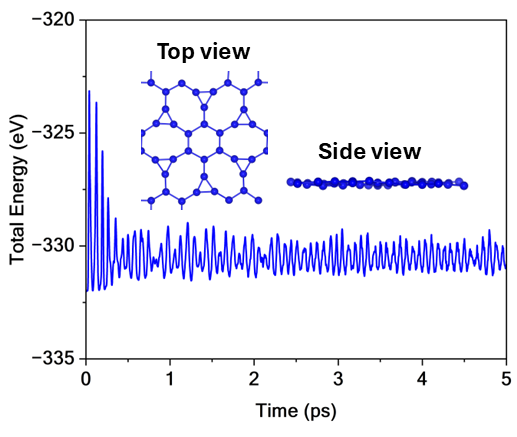}
\caption{Total energy profile of the $\beta$-IG monolayer during AIMD simulations at 300~K for 5~ps, along with top and side views of the final structure at the end of the simulation. The results indicate energy stabilization after initial fluctuations, confirming the structural integrity of the monolayer throughout the thermal process.}
\label{fig:aimd_pristine}
\end{figure}

To evaluate the thermal stability of the $\beta$-IG monolayer, AIMD simulations were carried out at 300 K. As shown in Figure~\ref{fig:aimd_pristine}, panel (a) presents the total energy evolution over a simulation time of 5~ps. After an initial fluctuation phase during equilibration, the energy stabilizes with small oscillations, suggesting that the system remains thermodynamically stable under ambient conditions. Panel (b) displays the top and side views of the final structure at the end of the simulation, revealing that the atomic framework of $\beta$-IG remains intact without bond breakage or significant structural deformations, further supporting its thermal robustness.

The elastic properties of $\beta$-IG were evaluated via in-plane elastic constants. As expected from its rectangular symmetry, the elastic tensor comprises four independent components: $C_{11} = 271.31$~N/m, $C_{22} = 260.12$~N/m, $C_{12} = 82.64$~N/m, and $C_{66} = 88.91$~N/m. The mechanical stability of 2D materials with orthorhombic symmetry can be verified by the Born-Huang criteria~\cite{born1954dynamical}, which require that $C_{11} > 0$, $C_{22} > 0$, $C_{66} > 0$, and $C_{11}C_{22} - C_{12}^2 > 0$. Therefore, the mechanical stability conditions are fulfilled, confirming the structural robustness of the monolayer under in-plane deformations.

\begin{figure*}[htbp]
\centering
\includegraphics[width=\linewidth]{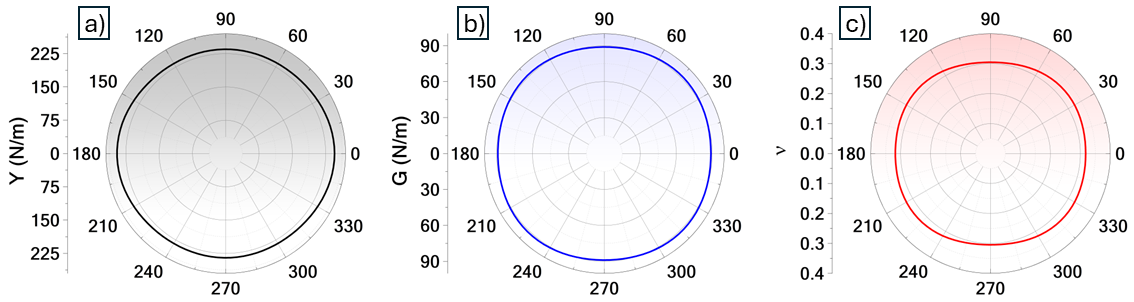}
\caption{(a) Polar representation of the orientation-dependent Young's modulus ($Y$), (b) shear modulus ($G$), and (c) Poisson's ratio ($\nu$) for $\beta$-IG. The near-circular patterns in all cases confirm the mechanical quasi-isotropy of the monolayer.}
\label{fig:mechprops}
\end{figure*}

The mechanical behavior of the $\beta$-IG monolayer was analyzed through the angular dependence of the in-plane elastic moduli, as shown in Figure~\ref{fig:mechprops}. The directional Young’s modulus ($Y$), shear modulus ($G$), and Poisson’s ratio ($\nu$) exhibit nearly isotropic behavior, with slight variations along different crystallographic orientations. Specifically, $Y$ ranges from 234.07 N/m to 245.06 N/m, and $G$ varies between 88.91 N/m and 91.49 N/m, 
indicating a stiff and moderately rigid monolayer. Poisson’s ratio also exhibits low directional variability, with values ranging from 0.305 to 0.325. These results reflect the high mechanical robustness and quasi-isotropic elastic response of $\beta$-IG, despite its rectangular symmetry.

\subsection{Electronic and Optical Properties}

\begin{figure}[htbp]
    \centering
    \includegraphics[width=0.75\linewidth]{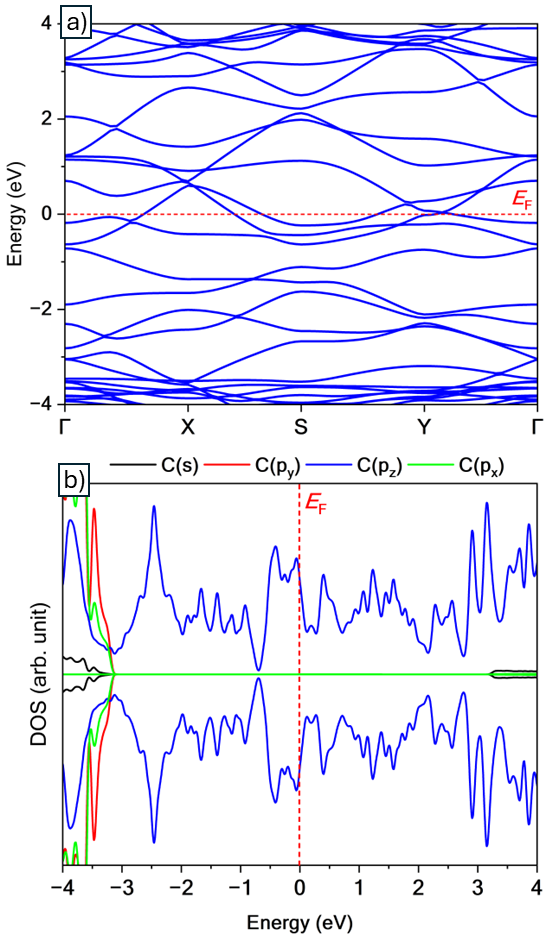}
    \caption{(a) Electronic band structure of pristine $\beta$-IG along the high-symmetry path in the Brillouin zone. The Fermi level ($E_F$) is set at 0~eV (dashed red line), and the bands crossing $E_F$ confirm the metallic character. (b) Total and orbital-resolved projected density of states (PDOS) showing the dominant contribution from C $2p_z$ orbitals near $E_F$, indicating significant $\pi$-electron delocalization.}
    \label{fig:bands_dos}
\end{figure}

The electronic band structure and the corresponding projected density of states (PDOS) of pristine $\beta$-IG are presented in Figure~\ref{fig:bands_dos}. The band diagram reveals the metallic nature of the system, with several bands crossing the Fermi level ($E_\mathrm{F}$) along the high-symmetry path $\Gamma$--X--S--Y--$\Gamma$. This feature is corroborated by the PDOS analysis, which shows substantial contributions near $E_\mathrm{F}$ from the carbon $p_\mathrm{z}$ orbitals—typical of $\pi$-bonded networks in $sp^2$-hybridized systems. The dominance of $p_\mathrm{z}$ states in the vicinity of $E_\mathrm{F}$ suggests strong electronic delocalization, which is favorable for enhancing in-plane electronic conductivity. These results indicate that $\beta$-IG possesses intrinsic conductivity.

\begin{figure}[htbp]
\centering
\includegraphics[width=0.75\linewidth]{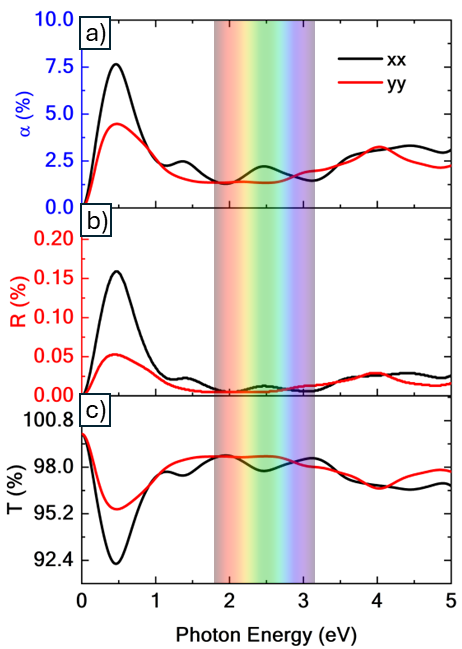}
\caption{$\beta$-IG optical characteristics: (a) absorption coefficient ($\alpha$), (b) reflectivity ($R$), and (c) transmittance ($T$) as functions of photon energy for a light polarized along the $x$ (black) and $y$ (red) directions. 
The shaded area highlights the visible spectrum (1.6--3.2~eV). The high transmittance and low reflectivity suggest excellent transparency and suitability for optoelectronic applications.}
\label{fig:optics}
\end{figure}

The $\beta$-IG optical properties were investigated through its absorption coefficient ($\alpha$), reflectivity ($R$), and transmittance ($T$), as shown in Figure~\ref{fig:optics}. All components were computed along both $x$ and $y$ directions to account for possible in-plane anisotropies. The absorption spectrum (Figure~\ref{fig:optics}a) exhibits distinct peaks below 2 eV and in the visible range, with a maximum intensity around 1.2 eV for the $x$-polarization, while the $y$ component is slightly weaker, indicating a small anisotropy. Reflectivity remains very low ($<0.2\%$) over the entire spectral range (Figure~\ref{fig:optics}b), favoring minimal optical loss. Remarkably, the transmittance exceeds 92\% in both directions (Figure~\ref{fig:optics}c), 
confirming that $\beta$-IG is a highly transparent monolayer, especially within the visible range (1.6--3.2 eV, shaded area). This high transparency, combined with moderate absorption, suggests potential applications in optoelectronic and transparent electrodes.

\subsection{Sodium Adsorption}

\begin{figure}
    \centering
    \includegraphics[width=0.9\linewidth]{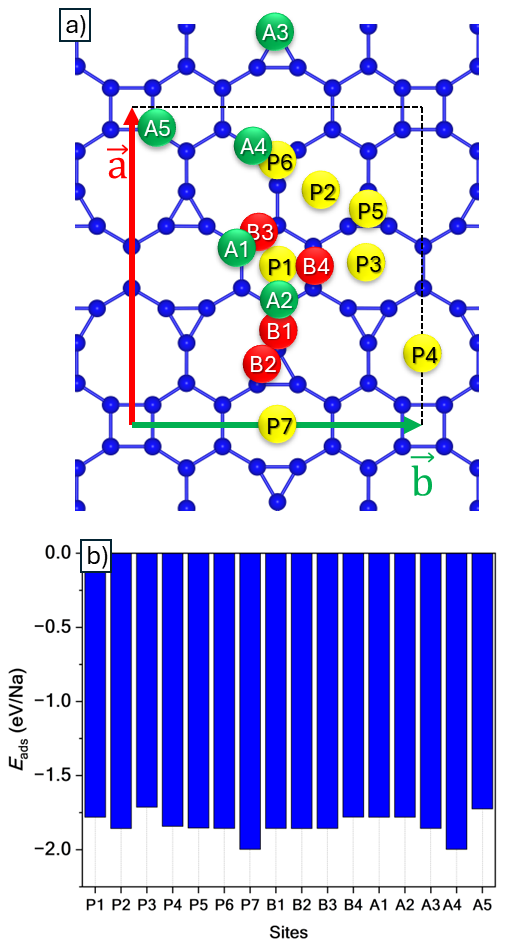}
    \caption{(a) Top view of the $\beta$-Irida-graphene monolayer showing all evaluated Na adsorption sites, labeled as P1--P7 (yellow), B1--B4 (red), and A1--A5 (green), respectively. The lattice vectors $\vec{a}$ and $\vec{b}$ are indicated in red and green, respectively. (b) Adsorption energy ($E_\mathrm{ads}$) for a single Na atom on each site indicated in Panel (a). Negative values indicate favorable binding configurations.}
    \label{fig:ads_sites}
\end{figure}

Figure~\ref{fig:ads_sites} illustrates the adsorption landscape of a single Na atom on the $\beta$-IG surface. Panel (a) shows the top view of the monolayer with candidate adsorption sites marked according to their local geometry. The sites labeled P1--P7 (yellow) correspond to hollow regions, while B1--B4 (red) are located above the bond centers, and A1--A5 (green) represent atop-like atom sites. 

Panel (b) shows the computed adsorption energy ($E_\mathrm{ads}$) for each site indicated in panel (a). All evaluated positions yield negative adsorption energies, confirming the energetic favorability of Na binding to the $\beta$-IG surface. Among them, site A4 and P7 exhibit the most favorable energies, approximately $-2.0$~eV, with the Na adatom positioned at P7 after optimization, indicating that sodium atoms preferentially occupy hollow regions above octagonal rings. 

\begin{figure}
    \centering
    \includegraphics[width=0.8\linewidth]{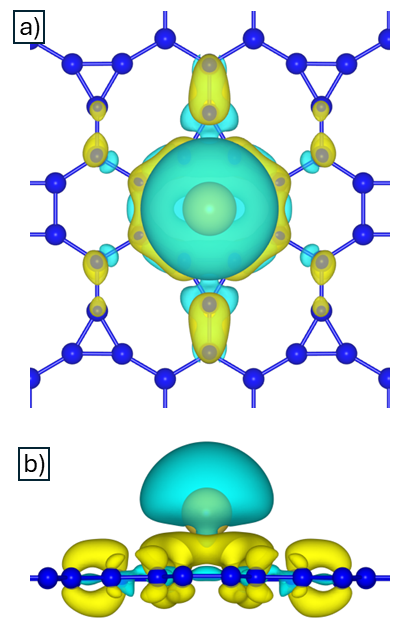}
    \caption{Charge density difference (CDD) isosurface for a Na atom adsorbed on P7 site of $\beta$-IG: (a) top view and (b) side view. Yellow and cyan regions represent charge accumulation and depletion, respectively, with an isovalue of 0.001 e/\AA$^3$. The plots clearly show significant charge transfer from Na to the $\beta$-IG monolayer, indicating ionic character of the interaction.}
    \label{fig:cdd}
\end{figure}

Figure~\ref{fig:cdd} displays the charge density difference (CDD) plots for a Na atom adsorbed at the P7 site of the $\beta$-IG monolayer: (a) top view and (b) side view. The CDD is defined as 

\begin{equation}
\Delta\rho = \rho_{\mathrm{Na@\beta\text{-}IG}} - \rho_{\mathrm{Na}} - \rho_{\beta\text{-}IG},
\end{equation}

\noindent where $\rho_{\mathrm{Na@\beta\text{-}IG}}$ represents the total charge density of the combined system, and $\rho_{\mathrm{Na}}$ and $\rho_{\mathrm{\beta\text{-}IG}}$ are the charge densities of the isolated Na atom and pristine $\beta$-IG, respectively, in the same geometrical configuration.

The yellow and cyan isosurfaces correspond to electron accumulation and depletion regions, respectively. Substantial electron accumulation is observed around the adsorption site on the $\beta$-IG surface, while the Na atom exhibits a pronounced electron depletion. This charge redistribution indicates a significant charge transfer from the Na atom to the $\beta$-IG sheet, consistent with the electropositive nature of Na. To further support this observation, Bader charge analysis was also performed, revealing a charge transfer of approximately 0.87$|e|$ from the Na adatom to the $\beta$-IG monolayer. This value indicates that the Na atom donates nearly its valence electrons, closely approaching the ideal oxidation state of +1. This substantial charge donation underscores the predominantly ionic character of the Na–$\beta$-IG interaction. Furthermore, the preservation of the planar structure of $\beta$-IG, as evident in the side view, confirms structural integrity upon Na adsorption.

\subsection{Anode Performance}

\begin{figure*}[htbp]
    \centering
    \includegraphics[width=0.75\linewidth]{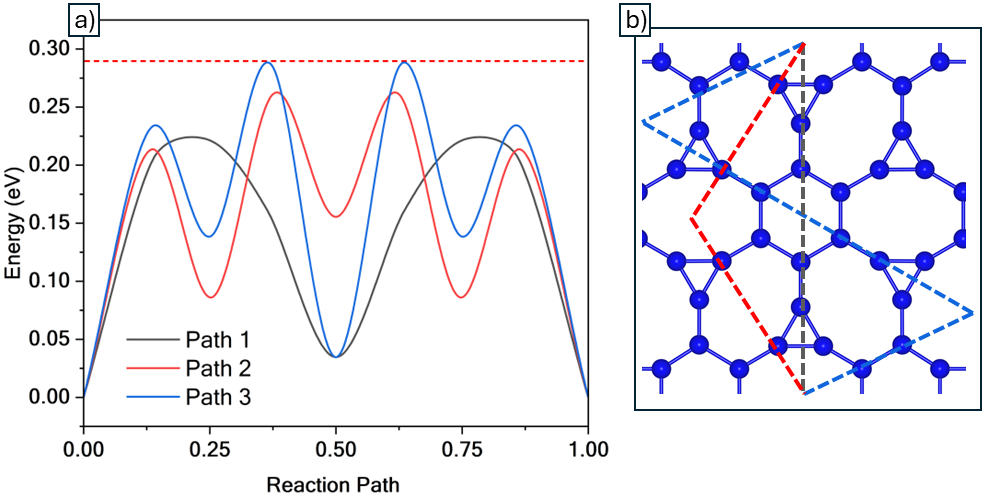}
    \caption{(a) CI-NEB energy profiles for Na diffusion along three representative paths on the $\beta$-IG surface. Path 2 shows the lowest energy barrier (0.16~eV), while Path 3 has the highest (0.27~eV). (b) Top view of the monolayer indicating the directions of the diffusion paths. The red dashed line corresponds to Path 2. The low diffusion barriers suggest high Na-ion mobility across the monolayer.}
    \label{fig:neb}
\end{figure*}

To evaluate the ionic mobility on the $\beta$-IG surface, the climbing-image nudged elastic band (CI-NEB) method was employed to calculate the energy barriers ($E_\mathrm{barr}$) for Na diffusion along representative pathways, as shown in Figure~\ref{fig:neb}. Three distinct diffusion paths were considered, corresponding to Na hopping between neighboring adsorption sites. The calculated diffusion barriers are 0.22 eV (Path 1), 0.26 eV (Path 2), and 0.29 eV (Path 3), all of which lie within a low-energy range conducive to fast charge/discharge rates in sodium-ion batteries. Notably, Path 1 presents the lowest barrier, indicating a favorable diffusion channel.

When compared to other 2D carbon-based systems, $\beta$-IG exhibits competitive or superior Na transport characteristics. For instance, graphene with grain boundaries exhibits energy barriers ranging from 0.09 to 0.35~eV, depending on the local environment~\cite{SUN2017415}. In C$_2$N monolayers and C$_2$N/graphene heterostructures, Li diffusion barriers fall within 0.2--0.5~eV~\cite{doi:10.1021/acs.jpcc.8b11044}, with sodiation being even less favorable due to larger ionic size (present case). In contrast, the C$_2$N-NHG framework supports efficient out-of-plane Na diffusion with a collective migration barrier as low as 0.18~eV~\cite{C9NR06011A}. Meanwhile, the carbon-rich C$_6$BN monolayer exhibits a Na diffusion barrier of 0.09~eV, one of the lowest reported~\cite{BUTT2025237119}. 

\begin{figure}
    \centering
    \includegraphics[width=1\linewidth]{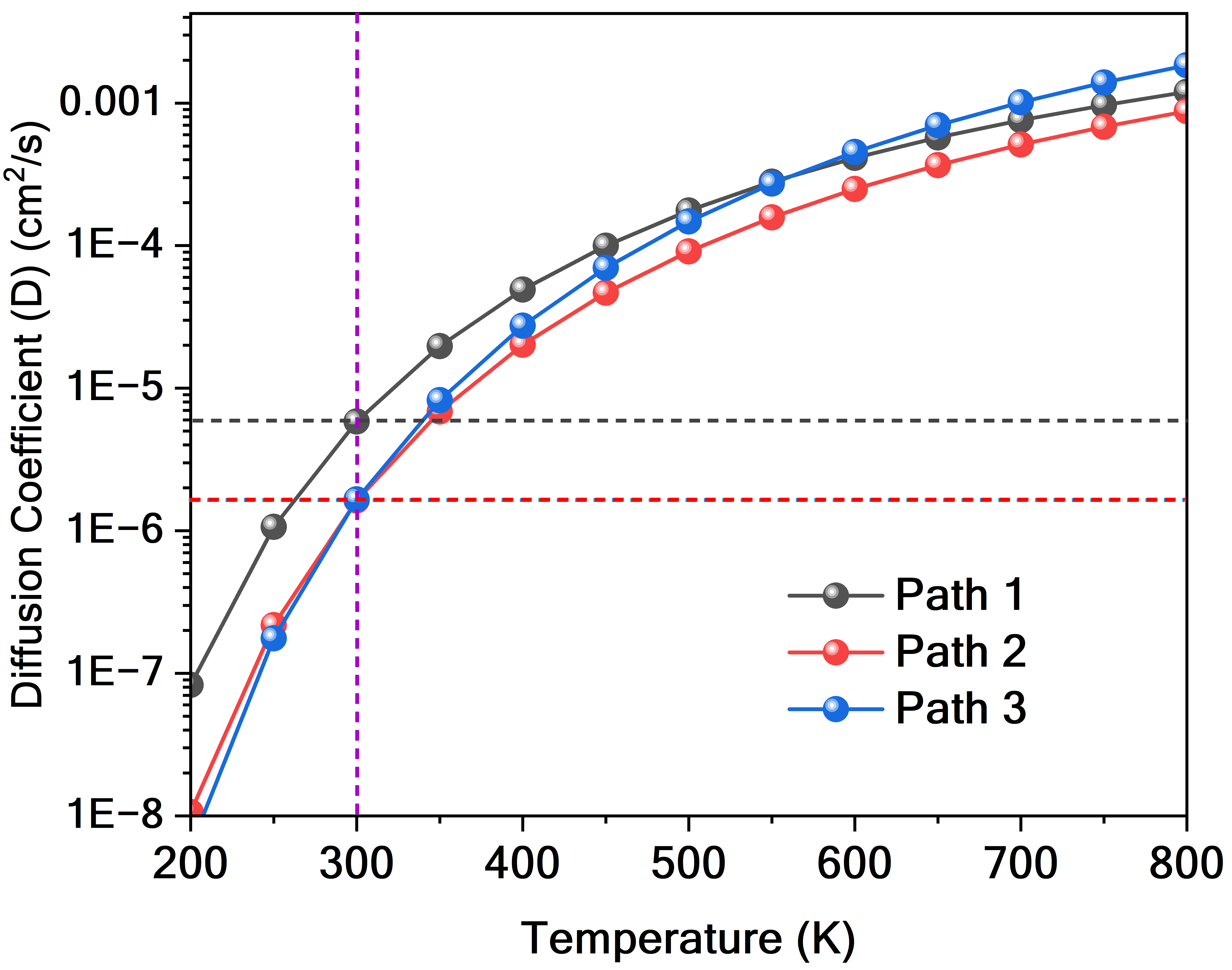}
    \caption{Diffusion coefficients ($D$) of Na atoms along the three identified diffusion pathways (Paths 1, 2, and 3) on $\beta$-IG as a function of temperature. The vertical dashed line marks 300~K. Horizontal dashed lines are visual guides for comparing $D$ values at room temperature.}
    \label{fig:diffusion}
\end{figure}

Figure~\ref{fig:diffusion} presents the temperature-dependent diffusion coefficients ($D$) of Na atoms for the three distinct migration pathways on $\beta$-IG, derived from the Arrhenius equation using CI-NEB-calculated activation energies. At near room temperature (300~K), highlighted by the vertical dashed line, Path 1 exhibits the highest diffusivity ($D \approx 5.85\times10^{-6}$~cm$^2$/s), followed by Path 3 ($\sim 1.66\times10^{-6}$~cm$^2$/s), and Path 2 ($\sim 1.63\times10^{-6}$~cm$^2$/s). All values are above the commonly referenced threshold of $10^{-6}$~cm$^2$/s for efficient Na-ion transport, with Paths 1 and 3 approaching or exceeding $10^{-6}$~cm$^2$/s. This high diffusivity indicates excellent Na mobility on $\beta$-IG, making it a compelling candidate for high-rate charge/discharge applications in SIBs.

Compared to other materials, $\beta$-IG has comparable diffusion characteristics. For instance, IG exhibits a diffusion coefficient of $3.11 \times 10^{-5}$~cm$^2$/s at 300~K, supported by a very low diffusion barrier of 0.09~eV~\cite{MARTINS2024114637}. HOP-graphene, another carbon allotrope, shows a Na diffusion coefficient of $2.78 \times 10^{-6}$~cm$^2$/s~\cite{MARTINS2025163737}. Similarly, Petal-graphyne achieves Li diffusivity around $10^{-8}$~cm$^2$/s~\cite{LIMA2025117235}. Monolayer pentagonal CoS$_2$, while not a carbon material, displays a comparable diffusion coefficient for Na ($3.13 \times 10^{-5}$~cm$^2$/s) with a slightly higher energy barrier of 0.22~eV~\cite{PhysRevApplied.16.024016}. Therefore, the Na transport behavior observed in $\beta$-IG, coupled with its structural stability and predicted capacity, highlights its good potential for advanced sodium-ion battery technologies.

\begin{figure}[htbp]
    \centering
    \includegraphics[width=1\linewidth]{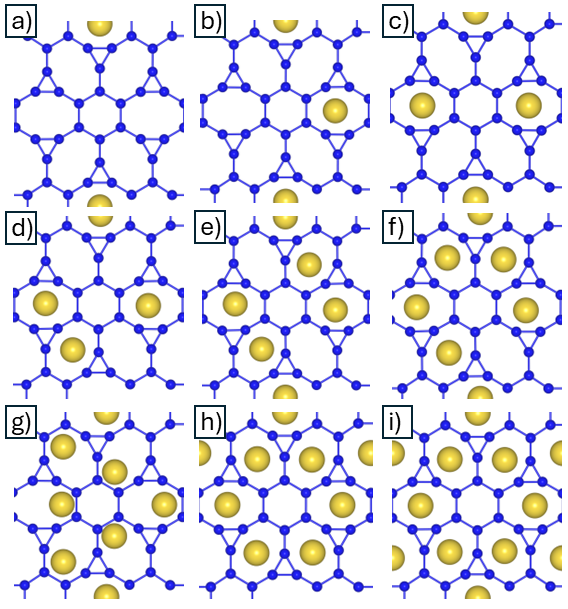}
    \caption{Top view of the optimized configurations for Na adsorption on $\beta$-IG, showing (a)--(i) increasing Na coverage with 2 to 18 atoms, adding Na atoms by pairs. The Na atoms (yellow spheres) are symmetrically placed above and below the monolayer, ensuring uniform charge distribution and mechanical stability.}
    \label{fig:na_adsorption_configurations}
\end{figure}

To investigate the Na storage capacity of $\beta$-IG, a series of configurations with an increasing number of adsorbed Na atoms symmetrically distributed on both sides of the monolayer was considered. Figure~\ref{fig:na_adsorption_configurations} illustrates the top view of the optimized geometries containing 2 to 18 Na atoms in the unit cell, which is composed of 38 carbon atoms. As Na coverage increases, the Na atoms remain uniformly distributed without significant distortion of the carbon backbone, indicating the structural robustness of the $\beta$-IG. This symmetric adsorption configuration is favorable for minimizing electrostatic repulsion and preserving structural planarity during ion adsorption.

\begin{figure}[htbp]
    \centering
    \includegraphics[width=0.85\linewidth]{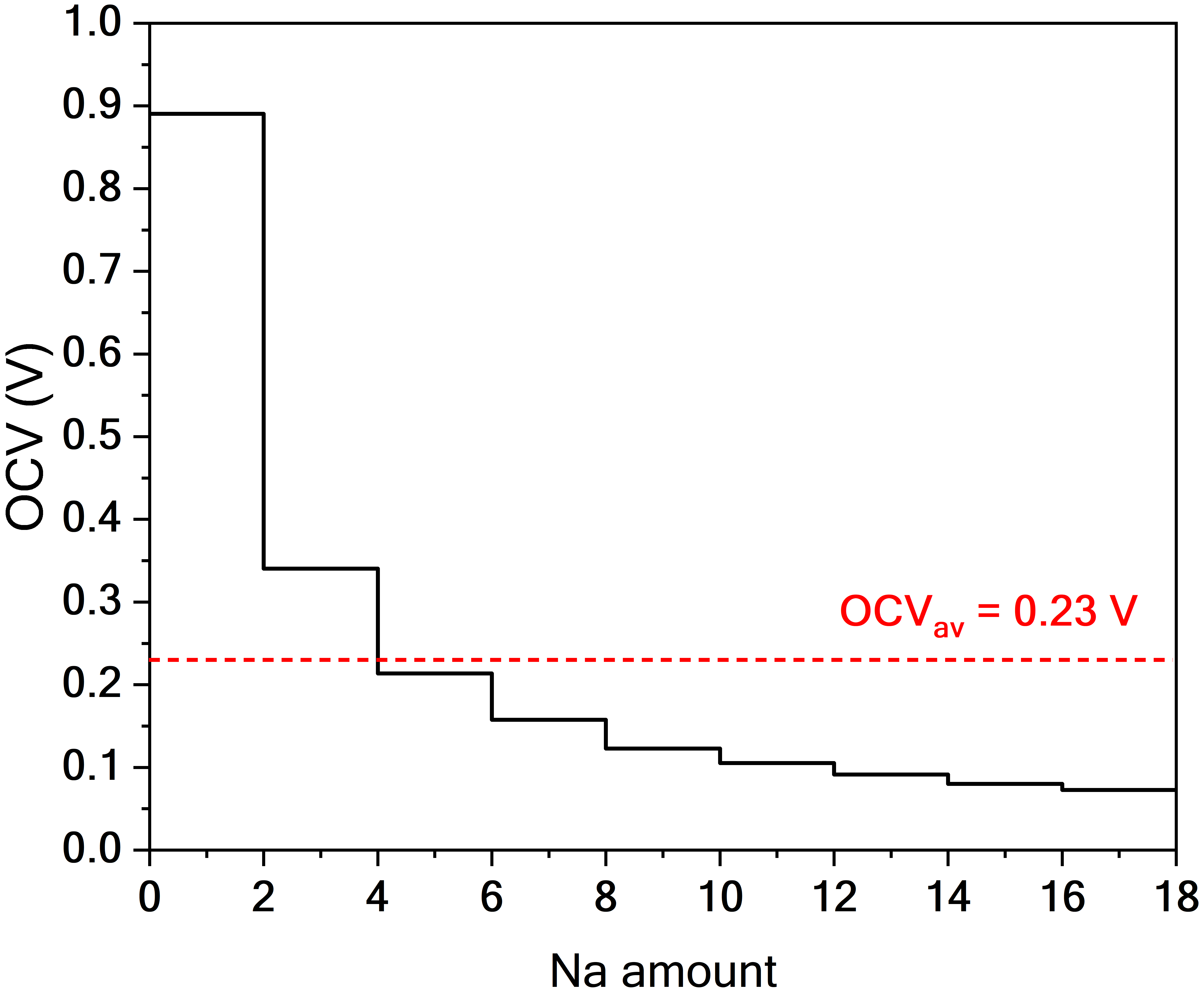}
    \caption{Open-circuit voltage (OCV) profile of $\beta$-IG as a function of the number of adsorbed Na atoms. The average voltage is ~ 0.23 V, indicated by the red dashed line.}
    \label{fig:ocv}
\end{figure}

Figure~\ref{fig:ocv} presents the open-circuit voltage (OCV) profile as a function of the number of adsorbed Na atoms on the $\beta$-IG monolayer. The OCV was calculated using the following relation~\cite{surila2025doping}:

\begin{equation}
    OCV = -\frac{E_{\mathrm{Na}_x\mathrm{C}} - E_{\mathrm{Na}_{x-1}\mathrm{C}} - E_{\mathrm{Na}}}{e},
    \label{eq:ocv}
\end{equation}

\noindent

where $E_{\mathrm{Na}_x\mathrm{C}}$ and $E_{\mathrm{Na}_{x-1}\mathrm{C}}$ represent the total energies of 
the $\beta$-IG monolayer with $x$ and $x-1$ adsorbed Na atoms, respectively, $E_{\mathrm{Na}}$ is the energy of a single Na atom in its bulk metallic state, and $e$ is the elementary charge. This formulation reflects the average intercalation voltage between successive Na loadings.

As shown in Figure~\ref{fig:ocv}, the initial Na insertions yield higher voltages, with a peak OCV close to 0.90 V. As the adsorption sites are occupied progressively, the voltage drops sharply and then stabilizes below 0.25 V after the fourth Na atom, eventually approaching 0.05 V at full Na coverage. The computed average OCV is 0.23~V, indicated by the dashed red line. This relatively low and stable voltage profile is advantageous for sodium-ion battery applications, offering high energy density while minimizing the risk of Na dendrite formation. Moreover, the absence of abrupt potential fluctuations supports a homogeneous and continuous Na adsorption process across the $\beta$-IG surface.

When compared to other 2D anode materials, $\beta$-IG exhibits a highly competitive electrochemical profile. For instance, the monolayer o-B$_2$N$_2$ presents an average Na voltage of 0.328 V\cite{KHOSSOSSI2022107066}, while the b-GeSe monolayer operates at an even lower voltage of 0.219 V\cite{zhou2018potential}. Likewise, the C$_2$N monolayer has been reported with an OCV of approximately 0.423 V\cite{Kadhim2023}, and AsC$_5$ monolayers deliver an average voltage of 0.17 V, with values ranging from 0.08 V to 0.32 V\cite{LU2023233439}. The V$_2$NS$_2$ monolayer exhibits Na OCV values ranging from 0.38 V to 0.86 V.

\begin{figure}[htbp]
    \centering
    \includegraphics[width=1\linewidth]{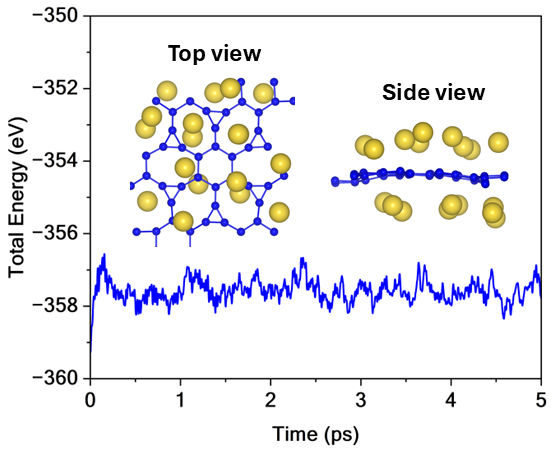}
    \caption{AIMD simulation of Na-decorated $\beta$-IG at 300~K for 5~ps. The total energy remains stable with only small fluctuations, indicating thermal robustness. Insets show the final configuration from the top and side views, where Na atoms (yellow) remain attached to the carbon sheet (blue) without clustering or detachment.}
    \label{fig:aimd}
\end{figure}

To further assess the thermal stability of the Na-decorated $\beta$-IG under ambient conditions, AIMD simulations at 300~K for 5~ps were performed. Figure~\ref{fig:aimd} displays the total energy fluctuations during the simulation. The low-amplitude oscillations and the absence of abrupt energy jumps indicate that the system remains structurally stable during the simulation. The insets present the final snapshots from top and side views, showing that the Na atoms remain well anchored to the surface without aggregation or detachment, thus confirming the thermal robustness of the complex.

The predicted specific capacity ($C_\mathrm{Q}$) of the $\mathrm{Na}_{18}\mathrm{C}_{38}$ system was estimated using the following relation~\cite{ullah2024theoretical}:
\begin{equation}
C_Q = \frac{nzF}{W} \times 1000,
\label{eq:CQ}
\end{equation}

\noindent where $n$ is the number of adsorbed Na atoms (18), $z$ is the valence state of Na ($z = 1$), $F$ is the Faraday constant (F = \SI{26.801}{Ah/mol}), and $W$ is the molecular weight of the sodiated structure. 

The calculated $C_\mathrm{Q}$ of $\beta$-IG is \SI{554.5}{mAh/g}, indicating its potential as a promising carbon-based 
anode for Na-ion batteries, as illustrated in Table~\ref{tab:comparison_capacity}. This value exceeds that of several well-established 2D materials. For example, the Ti$_2$B monolayer offers a Na storage capacity of 503.1 mAh/g~\cite{WANG2021148048},  while Ti$_3$C$_2$ MXene yields 351.8 mAh/g for Na~\cite{doi:10.1021/am501144q}. The Mn$_2$C monolayer achieves a similar performance (443.6 mAh/g), although with ultrafast diffusion~\cite{ZHANG2020144091}. Meanwhile, Mo$_2$C monolayer displays a relatively modest Na capacity of 132 mAh/g~\cite{doi:10.1021/acs.jpclett.6b00171}. Other materials, such as phosphorus carbide ($\alpha$-PC)~\cite{QI2019444} and MoN$_2$~\cite{C6TA07065E}, show Na capacities of 467.4 mAh/g and 864 mAh/g, respectively, although the latter includes significant contributions from transition-metal chemistry. In particular, $\beta$-IG surpasses conventional MXenes and carbon frameworks, offering an ideal balance between capacity and structural simplicity. These comparisons highlight the competitive edge of $\beta$-IG in terms of energy storage potential, particularly given its structural stability and low diffusion barriers.

\begin{table}[h!]
\centering
\caption{Comparison of predicted specific capacities ($C_Q$) for various 2D materials used as anodes for Na-ion batteries.}
\label{tab:comparison_capacity}
\begin{tabular}{lcc}
\hline
\textbf{Material} & \textbf{Na Capacity (mAh/g)} & \textbf{Reference} \\
\hline
$\beta$-IG & 554.5 & This work \\
Ti$_2$B  & 503.1 & \cite{WANG2021148048} \\
Ti$_3$C$_2$  & 351.8 & \cite{doi:10.1021/am501144q} \\
Mn$_2$C  & 443.6 & \cite{ZHANG2020144091} \\
Mo$_2$C  & 132.0 & \cite{ZHANG2020144091} \\
MoN$_2$  & 864.0 & \cite{C6TA07065E} \\
$\alpha$-PC  & 467.4 & \cite{QI2019444} \\
$\Theta$-graphene & 956.3 & \cite{WANG2019619} \\
\hline
\end{tabular}
\end{table}

\section{Conclusions}

This study provides an evaluation of the potential of the newly proposed $\beta$-Irida-graphene ($\beta$-IG) monolayer as a sodium-ion battery (SIB) anode material, utilizing density functional theory (DFT) simulations. The results demonstrate that $\beta$-IG exhibits remarkable thermal, dynamical, and mechanical stability, maintaining its structural integrity under ambient conditions, as confirmed by \textit{ab initio} molecular dynamics (AIMD) simulations and phonon dispersion analyses. The intrinsic metallic behavior of $\beta$-IG significantly facilitates efficient electronic transport, predominantly due to the $2p_z$-orbital contributions near the Fermi level. Sodium adsorption was energetically favorable, especially at hollow sites, with adsorption energies of approximately -2.0 eV. Furthermore, a substantial ionic character was evidenced by a Bader charge transfer of approximately $0.87|e|$ from sodium atoms to the $\beta$-IG monolayer. Calculated diffusion barriers ranging between $0.22$ and $0.29$~eV indicated excellent sodium-ion mobility, with diffusion coefficients as high as $5.85 \times 10^{-6}$~cm$^2$/s, enabling rapid charge and discharge cycles. Furthermore, $\beta$-IG achieved a competitive predicted specific capacity of $554.5$~mAh/g, surpassing or rivaling several advanced 2D materials reported recently in the literature. The stable and low-voltage profile, averaging around $0.23$~V, enhances the practical viability of $\beta$-IG by reducing the risk of dendrite formation. Collectively, these compelling attributes pose $\beta$-IG as an exceptional and structurally robust candidate, potentially advancing the development of sodium-ion battery anodes and broadening opportunities for future electrochemical energy storage technologies.

\section*{Data access statement}
Data supporting the results can be accessed by contacting the corresponding author.

\section*{Conflicts of interest}
The authors declare no conflict of interest.

\section*{Acknowledgements}
This work was supported by the Brazilian funding agencies Fundação de Amparo à Pesquisa do Estado de São Paulo - FAPESP (grant no. 2025/04757-6, 2022/03959-6, 2022/14576-0, 2020/01144-0, 2024/19996-3, 2024/19996-3, 2024/05087-1, and 2022/16509-9), and National Council for Scientific, Technological Development - CNPq (grant no. 307213/2021 – 8). L.A.R.J. acknowledges the financial support from FAP-DF grants $00193.00001808$ $/2022-71$ and $00193-00001857/2023-95$, FAPDF-PRONEM \\ grant $00193.00001247/2021 - 20$, PDPG-FAPDF-CAPES Centro-Oeste $00193-00000867/2024 - 94$, and CNPq grants $350176/2022 - 1$ and $167745/2023-9$. K.A.L.L. acknowledges the Center for Computational Engineering \& Sciences (CCES) at Unicamp for financial support through the FAPESP/CEPID Grant 2013/08293 - 7. Computational resources were provided by the Centro Nacional de Processamento de Alto Desempenho em São Paulo (CENAPAD-SP) and CENAPAD-RJ (SDumont). 

\printcredits

\bibliography{cas-refs}

\end{document}